\documentclass[twocolumn,preprintnumbers,amsmath,amssymb]{revtex4}

\usepackage{graphicx}
\usepackage{dcolumn}
\usepackage{bm}

\begin{document}

\title{The effects of solid-liquid interfacial tension on the settlement of sediment flocs}

\author{Zhao Jianglin}
\email{ydqc@eyou.com} \affiliation{Division of Engineering Science,
Institute of Mechanics, Chinese Academy of Sciences, Beijing,
100080, China}

\date{\today}

\begin{abstract}
In this paper, the effects of interfacial tension between the
sediment solid particle and  liquid on the settlement of sediment
flocs are investigated.  After a discussion of mechanical and
physical chemistry, we give a settling velocity expression including
such dynamical information of the floc growth as interfacial tension
and primary particle size \textit{etc.}. The resulting  expression
indicates the average settling velocity of sediment flocs increases
with increasing solid-liquid interfacial tension in a form of power
law and deceases with the primary particle size. We report on a
general method for analyzing settling behaviors of sediment flocs
under different flocculation conditions and verify the rationality
of the assumption of tension-induced flocculation by fitting typical
experimental data to the electrolyte concentration-dependent
sedimentation model which can follow from the relationship between
interfacial tension and electrolyte concentration.

\end{abstract}

\maketitle

\section{\label{sec:level1}  INTRODUCTION}

The aggregation-sedimentation of sediment flocs, which is widely
involved by many areas of industry and engineering such as mineral
processing, irrigation works, and hydraulic engineering, is a basic
phenomenon in the nature. Many fundamental researches have been
devoted to  understanding of it \cite{meh, bla,win1,win2,win3,all}.
Primary particles aggregate and form clusters or microflocs which
may continue to combine the remaining single particles or other
microflocs and thus become larger flocs. When the growth of flocs is
induced by the Brownian motion, the diffusion limited cluster
aggregation (DLCA) model or the reaction limited cluster aggregation
(RLCA) model can well describe the process of this aggregation.
However, when the flocs become large enough to settle under gravity,
sedimentation alters the growth mechanism. If the sediment
concentration $\phi$ is low, the flocs fall under gravity, grow,
fracturing and restructuring; the size of the flocs reach a final
value and the space-filling network connecting does not take place.
Conversely, if $\phi$ is sufficiently large, the growing microflocs
will form the space-filling networks which spans all the cell very
quickly.

The DLVO theory \cite{der,ver}, which attributes interactions
between colloidal suspensions to Van der Waals forces and
electrostatic repulsion, dynamically provides a quantitative
description of many facts of flocculation. In the DLVO framework,
colloidal suspensions tend to coagulate due to Van der Waals
attractive forces, while electrostatic repulsive forces promote
stability of the system, and the repulsive barrier may be decreased
through changing the physicochemical conditions, temperature, ionic
strength, PH, solvent quality and so on, which induces the
flocculation of colloidal particles. Although the DLVO theory can
describe basic facts of flocculation of colloids, it still can not
completely elucidate the cause of flocculation phenomena. For
example,  the DLVO theory  encounter many difficulties in
aggregation of charged colloidal particles in aqueous media such as
fluid-crystal coexistence phenomenon \cite{poo,kap,din}, and void
formation \cite{mov,tho}, etc, where a long-ranged attraction is
required. In addition, the DLVO theory is also reluctant to explain
the transition from RLCA to DLCA \cite{car}, since the fact that
high electrolyte concentrations lead to a fractal dimension arising
from DLCA model indicates that Van der Waals force cannot act as
such a role that may make formation of the floc attributed only to
pure Brownian collisions which means the binding force between flocs
should be independent of the floc itself.

To overcome discrepancies between theoretic predictions and
experimental observations, different versions of modification to the
DLVO theory, some of which have additionally considered non-DLVO
interactions resulting from acid-base \cite{van}, steric \cite{de},
and hydrodynamic \cite{gor} interfacial forces, have been presented.
However, most investigations into the effects of the interface  are
concentrated on interactions between particles \cite{kra,maz,fre}.
For a macroscopic system of sediment flocs, whose sizes can often
reach  500$\sim$1000 $\mu m$,  effects of interfacial tension
between the solid particle and liquid on mechanical balance of the
floc are more necessary to be taken into account, and on such a
large scale the DLVO force between particles becomes negligibly
small in comparison with gravity and hydrodynamic forces acting on
the floc, so that dynamical reason for sediment flocculation may be
attributed mainly to surface tension on the floc.

In this paper, we emphasize the effects of the interfacial tension
between solid particles and liquid on the settlement of sediment
flocs. Few works about settling dynamics of sediment flocs are
involved in the growth mechanism of flocs, and hence the aim of the
paper is to introduce such dynamical information as interfacial
tension, primary particle size (pps) \textit{etc.} in the settling
velocity expression of sediment flocs. We present a general method
for investigating sedimentation of sediment flocs under different
flocculation conditions and discuss settling behaviors influenced by
electrolyte concentration in relation with typical experimental
data.

\section{The settling velocity}
We now consider a system of sediment particles of the identical size
in still water, and focus on such the situation that the sediment
concentration $\phi$  is low so that spatially networking does not
happen. After a process of aggregating, large flocs fall under
gravity, then average settling velocity of the system is written as

\begin{equation}
U=\sum_i\psi_iv(D_i),
\end{equation}
where $\psi_i$ denotes the weight of the i-th group of flocs with
the same size $D_i$ , and $v(D_i)$ is (constant) settling velocity
of the floc belonging to the i-th group. It is necessarily
emphasized that $D_i$  should be understood as diameter of sphere of
the same volume as the floc including inside water. The isometric
diameter is preferentially considered here since some of experiments
of measuring settling velocity of sediment flocs are based on weight
change determination that directly connects with the isometric size.

Stokes settling velocity formula can be employed, if the floc are
not so large that effects of turbulence on it might be ignored, for
expressing the size-dependent sedimentation velocity $v(D_i)$ of the
floc \cite{jul}, namely

\begin{equation}
v(D_i)=\frac{1}{18}\frac{\rho_{Ai}-\rho_L}{\rho_L}g\frac{D_i^2}{\nu}=\frac{1}{18}\frac{(\rho_S-\rho_L)(1-\epsilon_i)}{\rho_L}g\frac{D_i^2}{\nu},
\end{equation}
where $\rho_{Ai,L}$ represents the density of the i-th floc and
liquid, respectively, $\nu$ is the viscosity of the liquid, $g$ is
acceleration of gravity, $\rho_S$ is density of the solid particle,
and $\epsilon_i$ denotes the porosity of the i-th floc. Here the
relation $\frac{\rho_{Ai}-\rho_L}{\rho_S-\rho_L}=1-\epsilon_i$ is
used. Generally, Eq. (2) is applicable for the case that the size of
the settling floc does not exceed 150 $\mu m$.

Investigations of floc structure \cite{lag,tam,mea} have shown that
the density (or porosity) of the floc may be evaluated from the
following relationship:
\begin{equation}
1-\epsilon=d_B(\frac{D_c}{D_0})^{-\alpha},
\end{equation}
where $d_B$ denotes a modifying factor, $D_c$ is the diameter of
collision, which denotes the size of the smallest sphere that fully
encompasses the floc, $D_0$ is the diameter of the primary particle
comprising the floc, and $\alpha$ is a constant. And the mass
fractal dimension $d_f$ can be scaled as $d_f=3-\alpha$. The number
$N$ of primary particles within a floc can be expressed in terms of
the collision diameter $D_c$ and the isometric size $D$,
respectively, namely
\begin{equation}
N=A_1(\frac{D_c}{D_0})^{d_f},
\end{equation}
where $A_1$ is the structure prefactor, and
\begin{equation}
N=(1-\epsilon)(\frac{D}{D_0})^3,
\end{equation}
as a result, the relation between $D_c$ and $D$  is given by
\begin{equation}
D_c=\sqrt[3]{\frac{d_B}{A_1}}D.
\end{equation}
Due to  Eq. (3) and Eq. (6), the sedimentation velocity $v(D_i)$ can
be rewritten as
\begin{equation}
v(D_i)=\frac{A_2}{18}\frac{(\rho_S-\rho_L)g}{\nu\rho_L}D_0^{3-d_f}D_i^{d_f-1},
\end{equation}
where $A_2=A_1^{\frac{3-d_f}{3}}d_B^{\frac{d_f}{3}}$. The
size-dependent velocity expression Eq. (7)  gives the information of
mass fractal dimension $d_f$, hence it is also be a theoretic
equation for measuring  $d_f$ of settling floc \cite{bus}.

\section{The additional pressure of surface of the floc}

Mechanism of growth of the floc affects sedimentation of it, hence
we have to take into account effects of the mechanism of the growth
on settling floc. It is pointed out in this paper that the
interfacial tension $\gamma$ on the solid-liquid interface (or the
Gibbs dividing surface) between sediment solid particle and liquid
plays a key role in aggregation and sedimentation of particles or
flocs. Obviously, the interfacial tension $\gamma$ balances with the
additional pressure difference $p$ between both sides of the
interface. The additional pressure difference $p$, which is
associated directly with stress which the floc is suffering,  is an
important quantity determining the structure and thus settlement of
the floc. For a regular surface,  $p$ is in inverse proportion to
radius of curvature of the surface, and then, for a floc surface
with the fractal feature, what properties does it have?

We assume, to expand the floc volume an increment $dV$, the
additional pressure needs to overcome the surface work of the
interface area adding $dS$ , that is,
\begin{equation}
p=\frac{\gamma dS}{dV}.
\end{equation}
and then from Eq. (5) we have
\begin{equation}
dS=\pi \beta_sD_0^2dN=(\pi\beta_s/D_0)d[(1-\epsilon)D^3];\ \ \
dV=\frac{1}{2}\pi D^2dD,
\end{equation}
where $\beta_s$ is the proportion factor denoting the ratio of solid
surface contributing to external surface of the floc to total solid
surface, which is associated with the shape of the floc. Combining
Eq. (3), (8), and (9), we derive the expression of the additional
pressure difference,
\begin{equation}
p=\frac{2\gamma d_f\beta_s A_2D_0^{2-d_f}}{D^{3-d_f}}.
\end{equation}
It is clearly seen that the additional pressure of the floc surface
is inversely proportional to the size of the floc since the fractal
dimension $d_f$ is generally less than 3, and it decays following
the power-law form, which is a reflection of abnormal specific
surface area of the floc, and also the additional pressure may
connect with electrolyte concentration through $\gamma$ according to
Gibbs adsorption law.

The reaction force of the additional pressure, acting on the floc,
provides a binding force that can stick particles together. However,
the inverse dependence on the floc size of the binding force
characterized by $p$ implies that not all positions on the surface
of a certain  floc can combine the other specific floc due to the
impact of external forces such as gravity, shear force,
\textit{etc.} on the specific floc. In other words, under the given
mechanical conditions, a floc would have a combinable region
relative to the other floc colliding with it, where both of them may
satisfy the mechanical condition of balance, as a result, a firm
agglomeration has to require many collisions of the two flocs, which
indicates a sticking probability less than 1 between flocs. In
particular, the sticking probability varies as different collision
pairs since it is associated with the ratio of the combinable region
area to the total surface area of either floc in a pair. Therefore,
interfacial tension-induced flocculation dynamics may result in a
growth mechanism ruled by the \textit{relative} sticking probability
during settling.


It is worthwhile to note that the pressure $p$ should be understood
as an average effect resulting from different solid surfaces
contributing to the external surface of the floc, which can be seen
from Eq. (8). In other words, for different solid surface elements
on the external surface of the floc, the pressure $p_i$ would be
fluctuating around the average $p$ deriving from Eq. (8).

\section{The maximal size of the floc }

The interfacial tension-induced mechanism of flocculation  implies
that there is a maximal diameter of the floc, since size of the floc
becoming large will abate the magnitude of the additional pressure
of the surface, which characterizes an ability of the floc's
capturing free particles or flocs. When sedimentation dominates over
Brownian motion, the floc sweeps up smaller flocs underneath and
grows faster, and then, when can the floc's size attain the maximum?

Let us assume that the floc is a sphere, then during settling, the
upper hemisphere will receive few smaller flocs due to gravity,
conversely the lower hemisphere can combine large numbers of smaller
flocs. And we find that the pressure required to equilibrate a
primary particle stuck on the floc, which is exerted by the
interface between the external surface of the floc and the liquid,
is minimum when the primary particle is enveloped on the bottom of
the sphere (See Fig. 1). Assuming the actual additional pressure,
thus the actual pressure acting on the primary particle, is
averagely the same down the external surface of the floc and is
determined by the size of the floc, we may know that the floc can
not stick other flocs together any more if it is not able to adhere
to the primary particle on its bottom. In other words, for the floc
of certain a size, the nearer the particle or  floc approaches the
bottom of the main floc, the easier it is fixed, hence shape of the
final floc generally is of cobble or strip. The case of the minimum
pressure, corresponding to that of the largest floc, is exactly what
we focus on.

If enough particles are supplied to make the floc largest, then for
a primary particle attached critically to the bottom of the floc,
mechanical equilibrium gives:
\begin{eqnarray}
p^{\prime}=-p, && G=p^{\prime}\Delta s,
\end{eqnarray}
where $p^{\prime}$ is the pressure acting on the primary particle
from the interface, $G$ is net weight of the primary particle in
water. Therefore the maximal diameter $D_m$ of the floc can be
derived from Eq. (10) and Eq. (11):
\begin{equation}
D_m=(\frac{\pi d_f \beta_s A_2\gamma}{c_B\Delta \rho
g})^{\frac{1}{3-d_f}}D_0^{\frac{1-d_f}{3-d_f}}.
\end{equation}
Here we apply $G=c_B(\rho_S-\rho_L)gD_0^3=c_B\Delta \rho gD_0^3$,
where $c_B$ is the shape factor,  and $\Delta s=\frac{1}{2}\pi
D_0^2$. It is clearly seen that the maximal size $D_m$ of the floc
increases in a form of power law as the interfacial tension $\gamma$
grows, while it drops with the size $D_0$ of the primary particle
since the fractal dimension $d_f$ is generally greater than 1 and
less than 3.

The expression of the maximal size Eq. (12) is a dynamical result
which excludes some geometrical factors such as competition between
the flocs in particles and spatial distribution of particles or
flocs. In other words, only if such dynamical parameters as
$\gamma$, $D_0$ \textit{etc.} can satisfy the critical condition of
balance, the maximal size of the floc may be uniquely determined.
However, actual average size of flocs in the system can not attain
the maximal size because of clustering competition, spatial
distribution, and sufficiency of particles. Therefore, the maximal
size $D_m$ is an imaginary quantity dependent only on dynamical
factors.

\begin{figure}
\scalebox{0.7}{\includegraphics{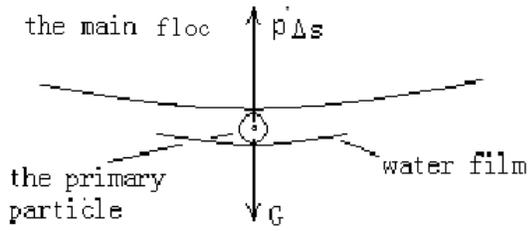}}
\caption{\label{fig:epsart} Schematic representation for mechanical
equilibrium under the situation of the minimum additional pressure,
thus the minimum $p^{\prime}$.}
\end{figure}

\section{The tension-dependent settling velocity}

In order to examine the effect of the interfacial tension on
settling velocity of sediment flocs, according to Eq. (1) and Eq.
(7), we write:
\begin{equation}
U=\frac{\lambda A_2}{18}\frac{\Delta \rho
g}{\nu\rho_L}D_0^{3-d_f}D_m^{d_f-1},
\end{equation}
where
$\lambda=\frac{\sum_i\psi_iD_i^{d_f-1}}{D_m^{d_f-1}}=\frac{\langle
D^{d_f-1}\rangle}{D_m^{d_f-1}}$.

We actually provide a  general method for investigating settling
velocity of sediment flocs, since the dimensionless number $\lambda$
may remove the effect of the dynamical factors such as the
interfacial tension and the primary particle size,  and is
associated only with geometrical factors like sediment concentration
and spatial distribution of particles, provided the growth of the
average size $\langle D^{d_f-1}\rangle$ indeed dynamically results
from those reasons which lead to the the maximal size of the floc.
In other words, the reason why the average size of flocs cannot
reach the maximal could be that multi-body competition in particles
under certain conditions reduces chances of the growth of each floc,
which exactly arises from geometrical factors.

Then the average velocity of  sediment flocs  can be expressed
through replacing $D_m$ in Eq. (13) with Eq. (12) as follows:
\begin{equation}
U=\frac{1}{18}\frac{\lambda}{\nu\rho_L } (\frac{\pi d_f
\beta_s\gamma}{c_B})^{\frac{d_f-1}{3-d_f}}A_2^{\frac{2}{3-d_f}}(\Delta
\rho g)^{\frac{2(d_f-2)}{d_f-3}}D_0^{\frac{4(2-d_f)}{3-d_f}}.
\end{equation}
We thus derive a setting velocity expression of sediment flocs
including dynamical information, which shows that the average
settling velocity of sediment flocs increases as solid-liquid
interfacial tension becomes large in a form of power law, and has
the decreasing relation with the primary particle size, since
Vladimir Nikora \textit{et al.} \cite{nik} have proved that the
fractal dimension $d_f$ can affect relationship between velocity and
size of the floc only when it is greater than 2. In addition,
because of high electrolyte concentrations resulting in larger
surface tensions, which is correct for inorganic salt solutes from
the knowledge of physical chemistry,  setting velocity would also
increase with the electrolyte concentration. We may also obtain the
expression of flocculation factor $F$:
\begin{equation}
F=\lambda (\frac{\pi d_f
\beta_s\gamma}{c_B})^{\frac{d_f-1}{3-d_f}}A_2^{\frac{2}{3-d_f}}(\Delta
\rho g)^{\frac{d_f-1}{d_f-3}}D_0^{\frac{2(1-d_f)}{3-d_f}},
\end{equation}
which is defined as $F=U/U_0$, where $U_0$ denotes the classical
Stokes¡¯ relationship for solid spherical particles, namely
$U_0=\frac{1}{18}\frac{\Delta\rho}{\rho_L}g\frac{D_0^2}{\nu}$. It is
obvious that the remarkable effect on flocculation does not lie in
electrolyte concentration but the size $D_0$ of  the primary
particle comprising sediment system, since the pps indicates the
essential factors affecting flocculation such as density of sediment
floc and binding force inducing flocculation \textit{etc.}, while
electrolyte concentration is only a reflection of influences of
solid adsorption action on flocculation. However, flocculation
factor expression Eq. (15) shows that $F$ varies inversely as
$\Delta \rho$ and $g$. It is not surprising since smaller $\Delta
\rho g$ will lead to larger average size of flocs when binding force
is fixed, yet this still needs further experimental proof.

\section{comparison with experimental data}

\subsection{The salinity-dependent settling velocity}

The effect of electrolyte concentrations on the settling behaviors
of sediment flocs may be derived from the relationship between the
surface tension and electrolyte concentration $C$. The relationship
between $\gamma$ and $C$ is exhibited in the Gibbs adsorption
equation:
\begin{equation}
\Gamma_{2,1}=-\frac{C_2}{RT}\frac{d\gamma}{dC_2},
\end{equation}
where $\Gamma_{2,1}$ is surface excess on the Gibbs' dividing
surface, $C_2$ is solute equilibrium concentration, $R$ is mole gas
constant and $T$ temperature. Therefore, if the isothermal
adsorption property $\Gamma_{2,1}(C_2)$ of the interface is given,
then the relationship between $\gamma$ and $C_2$  may be obtained
through integrating Eq. (16). The solid-liquid interface possesses
more complex adsorption properties and accordingly experimental
investigations on the relationship between $\Gamma$ and $C$ could be
more reliable to study the salinity-dependent settling behavior of
sediment flocs. Nevertheless, many physicochemical experiments of
sediment and soil show that the adsorption curve of some
solid-liquid interfaces can be expressed by Langmuir adsorption
isotherm, namely
\begin{equation}
\Gamma_{2,1}=b\frac{kC_2}{1+kC_2},
\end{equation}
where $b$ is the saturation adsorptive capacity, and $k$ is
adsorption constant.

However, for the interface between solid and liquid, we are more
interested in surface force relative to the solid rather than
relative to the bulk liquid. It is known that the addition of salt
to water will result in larger gas-liquid surface tension. This is
because attraction of the  salt ion in the bulk liquid for water
molecules can draw more water molecules inside the bulk liquid so
that more work would be needed to add a surface area, which shows
that the adsorption quantity of the dividing surface relative to the
bulk liquid is negative. In the same light, the adsorption of the
solid to inorganic ions will also give rise to a negative adsorption
quantity of the dividing surface relative to the bulk part between
the dividing surface and solid surface. Although the magnitude of
the adsorption relative respectively to two different bulk parts,
the bulk part between the dividing surface and solid surface, and
the bulk liquid, is the same (the dividing surfaces in the two cases
are different), the two quantities have opposite sign with each
other, and this distinction will play a key role in determining
properties of the interface tension.

Thus, combing Eq. (16) with Eq. (17) and noticing sign of the
adsorption quantity $\Gamma_{2,1}$, we get
\begin{equation}
d\gamma=\frac{bkRT}{1+kC_2}dC_2.
\end{equation}
After the integration of Eq. (18), a tension-concentration
relationship can be obtained:
\begin{equation}
\gamma=bRT\ln(1+kC_2)+\gamma_0.
\end{equation}
Here $\gamma_0$ denotes the interface tension between solid and pure
solvent. Finally, the salinity-dependent settling velocity $U$ and
flocculation factor $F$ can be expressed as follows,
\begin{widetext}
\begin{equation}
U=\frac{1}{18}\frac{\lambda}{\nu\rho_L } \{\frac{\pi d_f
\beta_s[bRT\ln(1+kC_2)+\gamma_0]}{c_B}\}^{\frac{d_f-1}{3-d_f}}A_2^{\frac{2}{3-d_f}}(\Delta
\rho g)^{\frac{2(d_f-2)}{d_f-3}}D_0^{\frac{4(2-d_f)}{3-d_f}}
\end{equation}
and
\begin{equation}
F=\lambda \{\frac{\pi d_f
\beta_s[bRT\ln(1+kC_2)+\gamma_0]}{c_B}\}^{\frac{d_f-1}{3-d_f}}A_2^{\frac{2}{3-d_f}}(\Delta
\rho g)^{\frac{d_f-1}{d_f-3}}D_0^{\frac{2(1-d_f)}{3-d_f}},
\end{equation}
\end{widetext}
respectively. It is apparent that both settling velocity of sediment
flocs and flocculation factor increase with increasing electrolyte
concentration. However, real adsorption of solid in the solution is
quite complex, even transitions between positive and negative
adsorption often occurs  for some cases. Therefore it would be more
reliable that experimental methods determine the relationship
between $\gamma$ and $C$ to predict the effects of electrolyte
concentration on settling velocity of sediment flocs. The
$\gamma$-$C$ form of Eq. (19) only reflects a simple case. In
addition, average size of flocs dependent on the maximal size $D_m$
can not be smaller than the pps $D_0$, hence the flocculation factor
$F$ is always greater than or equal to 1, which indicates a cutoff
concentration $C_{c}$ which guarantees $\langle D\rangle=\lambda
D_m\geq D_0$. The cutoff concentration $C_{c}$ actually is the
minimum salt concentration causing flocculation of particles of the
$D_0$ size in the present $\gamma$-$C$ form. If $\gamma$ is a
decreasing function of $C$, then $C_c$ which satisfies $F=1$ should
represent a minimum salt concentration giving rise to maximal
dispersion of flocs.


\subsection{The fitting of experimental data}

\begin{figure}
\scalebox{0.8}{\includegraphics{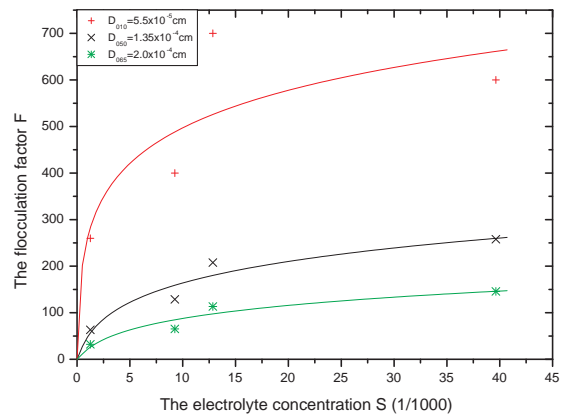}}
\caption{\label{fig:epsart} The fitting of experimental data to the
model Eq. (23). The experiments were performed with coastal silt at
the Bohai Bay in China of sediment concentration 4$\sim$5 kg/cm$^3$,
and the experimental temperature was about 293K. Solid lines denote
fitting curves and $D_{0**}$ is the cumulative average size of the
salt-free silt samples.}
\end{figure}

\begin{table*} \caption{\label{tab:table1}\textbf{Comparison
of Fitted Values of $F$ with Experimental Data as well as Values of
Related Parameters of Best Fit}}
\begin{ruledtabular}
\begin{tabular}{ccccccc}
 &\multicolumn{2}{c}{\ \ \ \ \ \ \ \ \ \ \ \ \ \ \ \ \ \ \ \ \ \ \ \ \ \ \ \ \ \ \ \ \ \ \ \ \ \ \ \
 \ \ \ \ \ \ \ \ \ \ \ \ \ \ \ \ \ \ \ \ \ \ \ $D_{065}=2\mu$m}\\
 Salinity&$F$\footnote{$F=\frac{U_{s65}}{U_{065}}, \frac{U_{s50}}{U_{050}}$, and $\frac{U_{s10}}{U_{010}}$,
 corresponding respectively to different $D_{0**}$s.}&$F$&\%Difference  &$B$(best fit)&$b$(best fit)&$k$(best fit) \\
$[1/1000]$& (experimental)&(fitted) & between the two   & & [Mol/cm$^2$]& [L/Mol]\\
 &&& $F$s &&& \\
\hline

 1.3&32.4&26.74 &17.5\%&$1.79\times10^{-21}$&0.99&62.49 \\
 9.3&65.4 & 85.04&23.1\%& & &  \\
 12.9&113.5&97.69 &13.9\%&\\
 39.7&146.0&146.18 &0.12\%& \\  \hline

 &\multicolumn{2}{c}{\ \ \ \ \ \ \ \ \ \ \ \ \ \ \ \ \ \ \ \ \ \ \ \ \ \ \ \ \ \ \ \ \ \ \ \ \ \ \ \
 \ \ \ \ \ \ \ \ \ \ \ \ \ \ \ \ \ \ \ \ \ \ \ $D_{050}=1.35\mu$m}\\

 1.3&63.4&57.62 &9.1\%&$8.42\times10^{-22}$&0.997&98.8 \\
 9.3&129.0&159.3 &19.02\%& & &  \\
 12.9&208.0&180.36 &13.3\%&\\
 39.7&258.0&259.89 &0.73\%& \\  \hline

 &\multicolumn{2}{c}{\ \ \ \ \ \ \ \ \ \ \ \ \ \ \ \ \ \ \ \ \ \ \ \ \ \ \ \ \ \ \ \ \ \ \ \ \ \ \ \
 \ \ \ \ \ \ \ \ \ \ \ \ \ \ \ \ \ \ \ \ \ \ \ $D_{010}=0.55\mu$m}\\

 1.3&260.0&285.42 &8.9\%&$7.53\times10^{-23}$&0.998&1634.75 \\
 9.3&400.0&488.93 &18.2\%& & &  \\
 12.9&700.0&526.29 &24.8\%&\\
 39.7&600.0&661.50 &9.3\%& \\

\end{tabular}
\end{ruledtabular}
\end{table*}

In order to examine the rationality of the assumption of
tension-induced flocculation, we fit the typical experimental data
which come from the report on sedimentation experiments of the
coastal silt at the Bohai Bay in China by Wu Deyi to the electrolyte
concentration-dependent sedimentation model Eq. (21) \cite{chi,
dey}. Wu Deyi \textit{et al.} utilize the sedimentation balance to
measure the electrolyte ($Na^+$) concentration-dependent settling
velocity in still water of silt flocs after analyzing the particle
size of salt-free silt sample by means of particle size analyzer.
The salt-free silt sample can be gained through washing silt with
distilled water and stirring and dispersing them in distilled water.
They report an empirical formula between settling velocity and the
salinity as well as the pps as follows,
\begin{equation}
\frac{U_s}{U_0}=h\cdot D_0^{-3.1}\cdot (S/1000)^{0.4},
\end{equation}
where $D_0$ is the average size of the salt-free silt sample,
$S/1000$ is the mass ratio concentration of $Na^+$, and the
numerical value of $h$ is estimated as 4$\times$10$^{-6}$. Due to
$F\sim D_0^{\frac{2(1-d_f)}{3-d_f}}$, we get the fractal dimension
$d_f$ in an average sense $d_f\simeq 2.2$ which is empirically
reasonable according to a large number of results of experimental
investigations on the fractal dimension of dispersions.

Let us rewrite Eq. (21) as
\begin{equation}
F=B\cdot[bRT\ln(1+k\frac{S}{23})+\gamma_0]^{\frac{d_f-1}{3-d_f}}(\Delta
\rho g)^{\frac{d_f-1}{d_f-3}}D_0^{\frac{2(1-d_f)}{3-d_f}},
\end{equation}
and take $R=8.31\times10^7$(dyne$\cdot$cm$\cdot$mol$^{-1}$K$^{-1}$),
$T=293$(K), $\gamma_0=72.92($dyne$\cdot$cm$^{-1}$), $\Delta
\rho=1.5$(g$\cdot$cm$^{-3}$), and $g=1000$(cm$\cdot$s$^{-2}$). Here
the unit of $k$ is [L/mol] and $B$ is a dimensionless parameter. We
use the least-squares fit of the model Eq. (23) to experimental data
reflecting the relationship between $F$ and $S$ to estimate values
of the parameters $B$, $b$, and $k$. The results are shown in Fig. 2
and Table. I.

It can be seen from the results of fit that the flocculation factors
obtained from the model Eq. (23) are close to those from real
experiments. Therefore the assumption of the tension-induced
flocculation of sediment particles are reasonable. The more
experimental data would attain better fits, nevertheless, percentage
difference between $F$s obtained by fits and experiments may still
provide a verification of the rationality of the model proposed. In
particular, the magnitude of the fitted parameter $k$ which
possesses definite physical sense accords with our understanding of
the adsorption of metal ions on sediment particles which is
generated from large numbers of experiments, for example, Ref
\cite{stu,smi,chen,gao,zin}, etc. In addition,
 the fact that values of the fitted parameter $b$ in
different measures  are approximately equal indicates that the
parameter $b$ may indeed represent such a physical quantity as
saturation adsorptive capacity.

\section{conclusion}

This paper proposes a theoretic model which attributes dynamical
reason of flocculation of flocs to the interfacial tension between
solid particles and liquid, and provides a general method of
analyzing sedimentation of sediment flocs. The resulting expression
 of settling velocity of sediment flocs can give a good description
 of the relationship between flocculation factor and the electrolyte concentration without loss of physical sense
 of fitted parameters. In fact, actual interaction is so  complex
 that more mechanical factors need to consider. For example, during
 aggregation ruled by Brownian motion while interacting flocs being effectively small,
 other forces have to be introduced into balance conditions. Further study should be
aimed at direct confirmation of the present model with real
experiments.

\end{document}